\documentclass[twocolumn]{aastex631}

\usepackage{graphicx}
\usepackage{amsmath}
\usepackage{color}
\usepackage{ulem}
\usepackage{float}
\usepackage{soul}
\usepackage{rotating}
\usepackage[graphicx]{realboxes}
\graphicspath{{fig/}}

\newcommand\lx{$L_{\rm 0.5 - 2\,keV}^{\rm gas}$}

\begin{document}

\title{Global Hot Gas Excess in (U)LIRGs: Replicating Galactic Nuclei Scaling Relations between Diffuse X-ray Emission and Star Formation on Galaxy-Wide Scales}
\correspondingauthor{Junfeng Wang}
\email{jfwang@xmu.edu.cn}

\author[0009-0007-7542-1140]{Chunyi Zhang}
\affiliation{Department of Astronomy, Xiamen University, 422 Siming South Road, Xiamen 361005, People's Republic of China}

\author[0000-0003-4874-0369]{Junfeng Wang}
\affiliation{Department of Astronomy, Xiamen University, 422 Siming South Road, Xiamen 361005, People's Republic of China}


\begin{abstract}

Hot ionized interstellar medium interlinks star formation and stellar feedback processes, redistributing energy, momentum, and material throughout galaxies. We use X-ray data from $Chandra$ to extract the hot gas emission from 78 of the most luminous infrared-selected galaxies in the local Universe. In the extreme star-forming environments, the intrinsic thermal X-ray luminosity of hot gas ({\lx}) shows a significant excess over the predictions of the standard linear $L_{\rm X}$$-$SFR relation for most objects with very high star formation rates (SFRs). The contribution of active galactic nuclei (AGNs) appears to have little impact on the global hot gas luminosity. For galaxies with SFR $\textgreater$ 50 ${M_{\rm \odot}}\,\,{\rm yr^{-1}}$, the Bayesian analysis gives a super-linear relation of ${\rm log}(L_{\rm 0.5-2\,keV}^{\rm gas} /{\rm erg\,s^{-1}})=1.34\,{\rm log}({\rm SFR}/{M_{\rm \odot}}\,{\rm yr^{-1}})+39.82$, similar to that found in the central regions of normal spiral galaxies. These results suggest a scenario in which the merger of galaxies delivers substantial amounts of gas, triggering intense star formation in both the nuclear region and the galactic disk, and ultimately enhancing the global thermal X-ray emission. The ratio of the apparent thermal luminosity in the 0.5$-$2 keV band ($L_{\rm 0.5 - 2\,keV}^{\rm appar}$) to {\lx} shows statistically significant negative correlations with the intrinsic column density ($N_{\rm H}$) and SFR. Moreover, in contrast to the luminosity ratio, SFR shows a moderate positive correlation with intrinsic $N_{\rm H}$. This suggests that the correlation between $L_{\rm 0.5 - 2\,keV}^{\rm appar}$/{\lx} and SFR may be driven by the underlying $L_{\rm 0.5 - 2\,keV}^{\rm appar}$/{\lx}$-$$N_{\rm H}$ and SFR$-$$N_{\rm H}$ relations.

\end{abstract}


\section{Introduction}\label{sec:intro}

Luminous infrared galaxies (LIRGs) are systems with infrared luminosities exceeding $10^{11}\,L_{\rm \odot}$, encompassing a diverse population that ranges from isolated galaxies to interacting pairs and late-stage mergers. LIRGs are typically gas-rich, which fuels intense star formation and enhanced AGN activity \citep{1989_Hernquist_Natur_687H, 2005_Matteo_Natur_604D}. Moreover, in the most luminous regime, ultraluminous infrared galaxies (ULIRGs; $L_{\rm IR}$ $\geqslant$ $10^{12}\,L_{\rm \odot}$) are considered an important evolutionary stage toward the formation of quasars (QSOs) \citep{1988_Sanders_ApJ_74S,1988_Sanders_ApJ_35S,2005_Hopkins_ApJ_705H} and massive elliptical galaxies \citep{1972_Toomre_ApJ_623T, 2002_Tacconi_ApJ_73T, 2006_Cox_ApJ_791C,2017_Sparre_MNRAS_3946S}.

These objects contribute significantly to the cosmic infrared background and are common at redshifts 1$-$3, where the peak of star formation occurs in the Universe \citep{2014_Casey_PhR_45C}. Accordingly, they certainly play a key role in deepening our understanding of the general evolution of galaxies and black holes.

In the scenario proposed by \citet{1988_Sanders_ApJ_74S} and \citet{2005_Hopkins_ApJ_705H}, ULIRGs at a late stage of merger are expected to consume or expel their gas and subsequently evolve into an obscured Type II QSO, and eventually become an exposed (Type I) QSO. This process ultimately leads to the formation of an elliptical galaxy and contributes significantly to the growth of the central supermassive black hole. However, more recent simulations indicate that when mergers preserve substantial reservoirs of cold interstellar gas, this material can subsequently settle into a rotationally supported plane and form a new disk \citep[e.g.,][]{2002_Barnes_MNRAS_481B, 2005_Springel_ApJ_9S, 2006_Robertson_ApJ_986R}. This means that the final galactic morphology (elliptical or spiral) following a merger event critically depends on the evolution of the interstellar gas. 

The amount of the interstellar gas is regulated by the star formation processes and AGN, both of which inject energy into the surrounding medium. In particular, stellar winds and supernova feedback from massive stars heat the gas intensely and can even drive it into the galactic halo \citep{2005_Grimes_ApJ_187G, 2012_Mineo_hotgas, 2013_LiJiangTao_MNRAS_2, 2013_Hopkins_MNRAS_78H}. AGN feedback also heats the interstellar gas and has the potential to quench star formation by preventing gas cooling and collapse \citep{2005_Matteo_Natur_604D, 2005_Springel_ApJ_9S, 2015_Choi_MNRAS_4105C}.

X-ray observations are a powerful tool for studying these high-energy feedback phenomena. In spiral galaxies, the thermal diffuse X-ray emission from hot gas shows a strong linear correlation with the star formation rate \citep{2012_Mineo_hotgas}. The ratio $L_{\mathrm{X}}(\mathrm{gas})/{\rm SFR} \thickapprox 5.5 \times 10^{39} (\mathrm{erg \, s^{-1}})/({\rm M_{\odot}\ yr^{-1}})$ from \citet{2018_Smith_AJ_81S} indicates that the feedback in star-forming galaxies reaches a quasi-steady state, where about 2\% of the mechanical energy from supernovae and stellar winds is converted into X-ray emission. However, unlike the linear $L_{\rm X}$$-$SFR scaling relation, a recent study based on the first eROSITA all-sky survey found a significant excess in the integrated X-ray luminosity at low SFRs ($\textless$ 1 ${M_{\rm \odot}}\,\,{\rm yr^{-1}}$) compared to the expectations from the linear scaling relation \citep[e.g.,][]{2025_Kyritsis_A&A_128K}. This integrated result implies that in galaxies with low-level star formation, many of which are very faint or distant, the properties of the hot ionized interstellar gas emission may differ from our anticipations. Nevertheless, for extremely high SFR galaxies, it remains unclear whether the hot gas emission follows the standard linear $L_{\rm X}$$-$SFR scaling relation.

In this paper, we aim to investigate the hot gas $L_{\rm X}$$-$SFR relation in nearby LIRGs and ULIRGs, and to examine whether the X-ray excess exists in these extreme star-forming galaxies. We apply a dust extinction method to estimate the intrinsic hydrogen column density ($N_{\rm H}$) for 78 (U)LIRGs selected from the C-GOALS \citep[the X-ray component of the GOALS multi-wavelength survey;][]{2011_Iwasawa_A&A_106I,2018_Torres_A&A_140T} program, and then correct the observed diffuse X-ray emission for intrinsic absorption.

This paper is organized as follows: In Section~\ref{sec:data} we describe the sample of galaxies, the data, and the methods used in our data reduction and analysis. In Section~\ref{sec:results} we present our main results on the properties of hot gas in (U)LIRGs. Conclusions are summarized in Section~\ref{sec:summary}. 

\section{data and reduction} \label{sec:data}

\subsection{Sample}\label{sec:Sample}

Based on the Great Observatories All-sky LIRG Survey \citep[GOALS\footnote{\url{http://goals.ipac.caltech.edu/}};][]{2009_Armus_PASP_559A}, we selected 78 (ultra)luminous infrared galaxies with $Chandra$ X-ray data of adequate quality to study the diffuse X-ray-emitting gas from C-GOALS \citep[$Chandra$-GOALS;][]{2011_Iwasawa_A&A_106I, 2018_Torres_A&A_140T}. GOALS is a comprehensive, multi-wavelength study of the brightest infrared galaxies in the local Universe, with targets selected from the IRAS Revised Bright Galaxy Sample \citep[RBGS;][]{2003_Sanders_AJ_1607S}. The RBGS comprises 629 extragalactic objects detected by IRAS, all with 60 $\mu$m flux density greater than 5.24 Jy, and provides full-sky coverage above Galactic latitude $|b|$ $\textgreater$ $5^\circ$. As the brightest infrared galaxies in the sky, these objects are unparalleled laboratories for studying hot gas in extreme star-forming environments.

\begin{figure*}
\centering 
\includegraphics[scale=1.0]{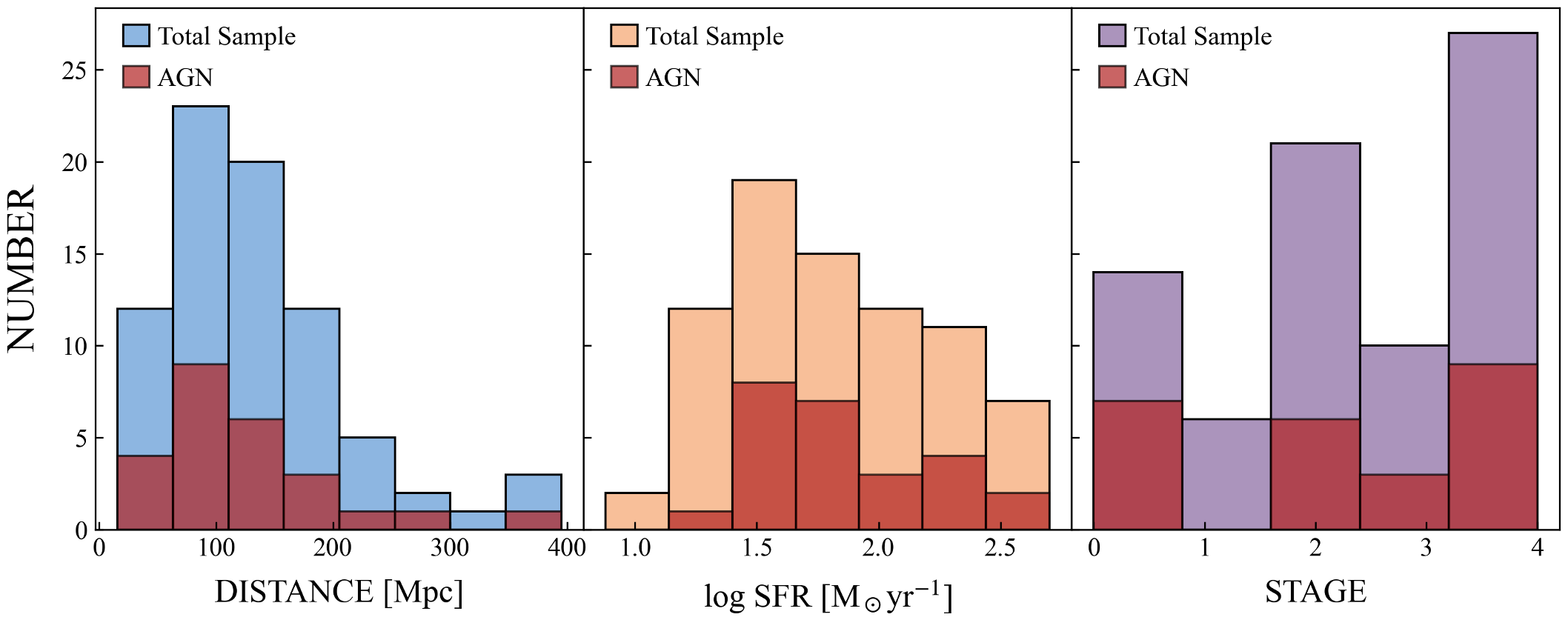}
\caption{Distribution of distance, SFR, and merger stage for the sample galaxies. The numbers of AGN are marked in red in each bin. More than 80\% of the galaxies are in merger systems, while this fraction reaches 100\% for ULIRGs. }
\vspace{0.4cm}
\label{fig:sample_statistic}
\end{figure*}

The sample in this work consists of 14 isolated galaxies and 58 systems in different merger stages. A total of 12 galaxies in the 6 systems (NGC 6670, NGC 5331, KTS 57, KTG 82, NGC 7592, and Arp 293) were resolved in both GALEX FUV and point source-excluded X-ray images. Table~\ref{tab:table1} presents the basic parameters for all 78 galaxies in the C-GOALS sample. The redshift range of the sample galaxies is $z$ = 0.003$-$0.0857, with total SFR spanning $\thicksim$2 dex (e.g., $\thicksim$6$-$501 ${M_{\rm \odot}}\,\,{\rm yr^{-1}}$) at the extreme high-SFR end. The distributions of distance, SFR, and merger stage are shown in Figure~\ref{fig:sample_statistic}.

\subsection{Extraction of X-Ray Spectra }\label{sec:Extraction}

All of the $Chandra$ data used in this study were downloaded from the $Chandra$ Data Archive. The detailed observation ID (obsid) and exposure time for each galaxy are listed in Table~\ref{tab:table2}. We reprocessed these data using the $Chandra$ Interactive Analysis of Observations (CIAO) software version 4.17 with CALDB version 4.11.6. When multiple observations were available for an object, we combined them to enhance the sensitivity.

The point spread function (PSF) of $Chandra$ at each sky location was determined using the CIAO routine $mkpsfmap$. With the PSF map and the wavelet-based source detection algorithm $wavdetect$ \citep{2002_Freeman_ApJS_185F}, the point source detections were performed on the scales of 1, $\sqrt{2}$, 2, 2$\sqrt{2}$, 4, and 8 pixels ($0^{\prime \prime}.492$/pixel) and in the soft (0.5 ${-}$ 2.0 keV), hard (2.0 ${-}$ 8.0 keV), and total (0.5 ${-}$ 8.0 keV) energy bands. The radius of the point source was set to enclose 90\% of the source counts based on the PSF shape. For brighter sources, we manually increased the size of the radius to remove visible ring-like features due to spillage from the PSF wings \citep{2004_Gaetz_SPIE_411G,2004_Allen_SPIE_423A}. The point source removal was performed individually for each observation as the PSF of an object could vary significantly for multiple observations.

The CIAO routine $specextract$ was used to extract the point source-excluded spectra to obtain the diffuse X-ray emission in the 0.5$-$2 keV band. The background regions were selected from source-free areas on the same CCD chip. For the majority of the sample galaxies, which have few counts, we grouped the spectra to achieve a signal-to-noise ratio (S/N) $\textgreater$ 2 per bin. Thus, we used the C statistic \citep{1979_Cash_ApJ_939C} as the fit statistic instead of chi-squared.

\subsection{Correction for Intrinsic Absorption}\label{sec:Absorption}

To determine the intrinsic X-ray luminosity of the hot gas, it is necessary to take into account the absorption of the foreground gas in the Milky Way and the gas within the host galaxy. For Galactic absorption, we used the Galactic hydrogen column densities $N_{\rm H}$ from the LAB HI map drawn by \citet{2005_Kalberla_A&A_775K}. The values of Galactic $N_{\rm H}$ are shown in Table~\ref{tab:table1}. For intrinsic absorption of galaxies, the high S/N X-ray data of luminous targets could constrain the intrinsic hydrogen column density through the X-ray spectra.

However, the intrinsic $N_{\rm H}$ cannot be reliably constrained by the X-ray spectra for most targets due to large uncertainties. Thus, we use dust extinction to estimate the intrinsic $N_{\rm H}$ (e.g., the dust-to-gas ratio) in this work. Following \citet{2011_Hao_ApJ_124H}, we utilized the total infrared to FUV luminosity ratio (IRX) to determine the FUV extinction $A_{\rm FUV}$. We then converted $A_{\rm FUV}$ to optical extinction $A_{\rm V}$ using \citet{2001_Calzetti_PASP_1449C} dust attenuation law. Combining the extinction curve of $A_{\rm V}$/$E$($B$$-$$V$) = 3.1 \citep{1979_Savage_ARA&A_73S} and $N_{\rm H}$/$E$($B$$-$$V$) = 8.8 $\times$ $10^{21}$ cm$^{-2}$ mag$^{-1}$ \citep{2017_Lenz_ApJ_38L}, we inferred the color excess $E$($B$$-$$V$) and calculated the intrinsic $N_{\rm H}$ for each sample galaxy. The derived values of the intrinsic $N_{\rm H}$ are listed in Table~\ref{tab:table3}.

Although the method of dust-to-gas ratio can estimate the hydrogen column density, we note that it may not always be reliable since this method depends on the assumed geometry of the hot gas relative to the young massive stars. Therefore, in order to check on the intrinsic $N_{\rm H}$ derived from $A_{\rm V}$, we modeled 11 X-ray bright objects, whose spectra contain at least 20 bins with the S/N $\textgreater$ 5$\sigma$ per bin, using the model described in Section~\ref{sec:Fitting}.

\begin{figure}
\includegraphics[scale=0.55]{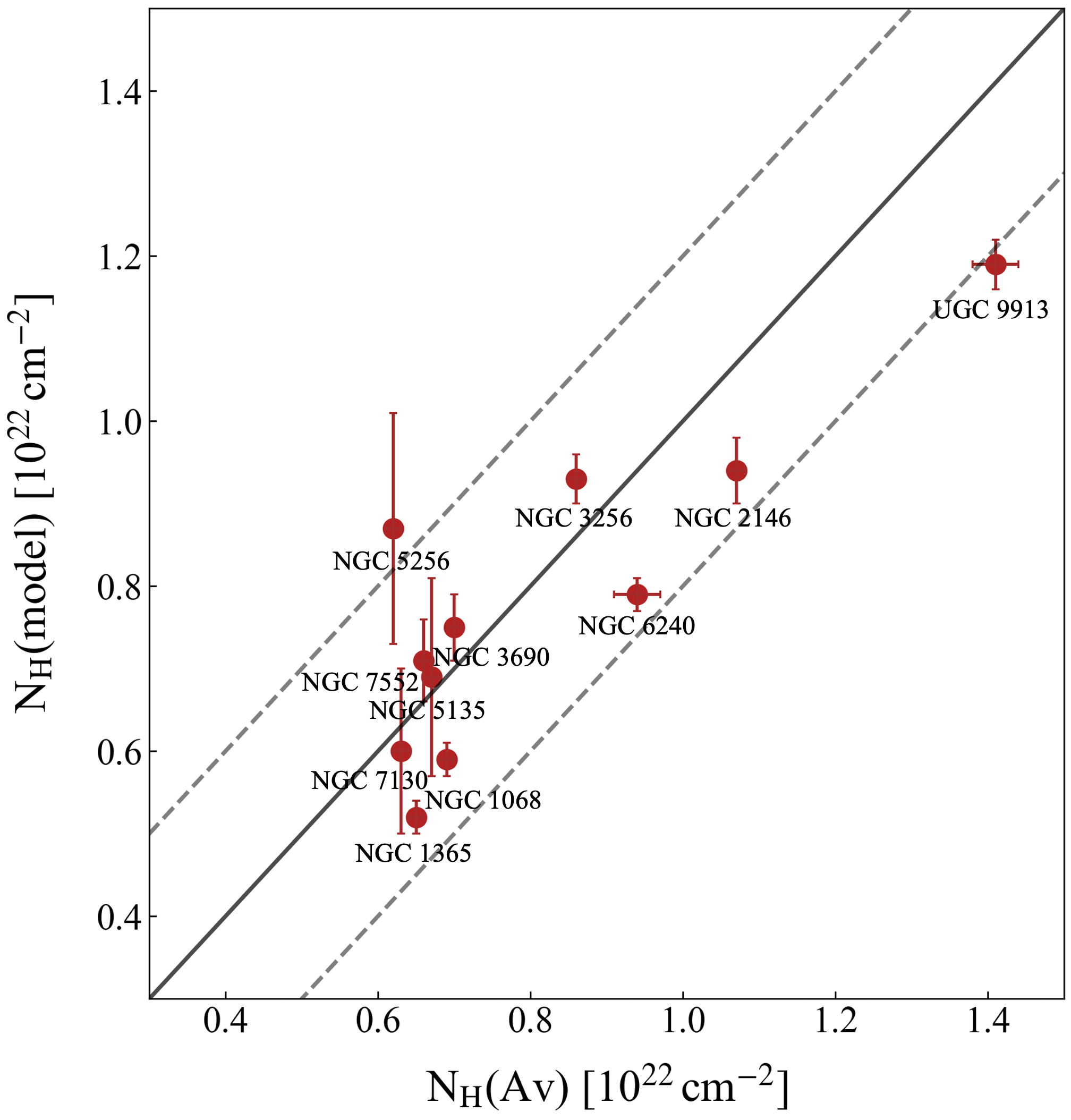}
\caption{Comparison of intrinsic $N_{\rm H}$ estimated from dust extinction (e.g., the dust-to-gas ratio) and from best-fitting models of X-ray spectra for 11 sample galaxies. The solid line corresponds to the $y=x$ relation, and the dashed lines represent $y=x \pm 0.2$. }
\vspace{0.1cm}
\label{fig:Av-model}
\end{figure}

Figure~\ref{fig:Av-model} shows the intrinsic $N_{\rm H}$ given by the best-fitting model compared to the $A_{\rm V}$-derived $N_{\rm H}$. For 9 out of 11 objects, the two methods yield consistent $N_{\rm H}$ with differences of less than 0.2 $\times$ 10$^{22}$ cm$^{-2}$, and the remaining two (NGC 5256 and UGC 9913) are slightly above this range. We have calculated the X-ray luminosities corrected using both methods and found no statistically significant differences. Thus, the $A_{\rm V}$-derived $N_{\rm H}$ can serve as the absorbing column density of the host galaxy when more direct X-ray spectral measurements are unavailable.

\subsection{Spectral Fitting}\label{sec:Fitting}

We used the thermal plasma emission model APEC \citep{2001_Smith_ApJ_91S} and the power-law model to fit the point source-excluded X-ray spectra in the xspec\footnote{\url{https://heasarc.gsfc.nasa.gov/xanadu/xspec/}} software version 12.14.1. In star-forming galaxies, the hot gas is primarily generated and heated through supernova shocks and stellar winds from massive stars \citep{2005_Grimes_ApJ_187G, 2019_Fabbiano_cxro.book_7F, 2022_Nardini_hxga, 2025_Zhangcy_ApJ_15Z}. Given that the APEC component is generally considered to be dominated by the emission from hot gas, throughout this work, we assume this plasma component originates from hot gas and refer to the absorption-corrected luminosity of the APEC component in the 0.5$-$2 keV band as {\lx}. The power-law component is the light from faint unresolved compact sources, which are below the point source detection threshold and typically have a non-thermal spectrum.

For our initial fits, we varied the plasma temperature T and the power-law photon index $\Gamma$ of the spectral model consisting of an absorbed APEC component plus an absorbed power-law, which we refer to hereafter as model (1). This fit is generally good for the sample galaxies. However, to obtain the best-fitting model for the X-ray spectra, we ran another six spectral models, including (2) fixing $\Gamma$ = 1.8 and T = 0.24 keV for the absorbed power-law and APEC components, (3) fixing $\Gamma$ = 1.8 and T = 0.71 keV, (4) fixing $\Gamma$ = 1.8, (5) fixing T = 0.24 keV, (6) fixing T = 0.71 keV, (7) two absorbed APEC components plus the power-law with both T and $\Gamma$ free. For these models, $\Gamma$ = 1.8 is typically expected for X-ray binaries and AGNs in nearby galaxies \citep{1994_Nandra_MNRAS_405N,2002_Kong_ApJ_738K,2004_Swartz_ApJS_519S}. T = 0.24 keV and 0.71 keV are the characteristic temperatures of hot gas in spiral galaxies \citep{2005_Grimes_ApJ_187G, 2012_Mineo_hotgas}.   

We compared the degrees of freedom and the C-statistics of these models to identify the good-fitting models. We rejected models that yielded non-physical results, such as very high/low temperatures or extreme photon indices. When multiple good models with different parameters were available, we preferred the one with a higher plasma temperature and fewer degrees of freedom. 

Table~\ref{tab:table3} presents the best-fitting models for all objects, along with the thermal X-ray luminosities corrected for the intrinsic absorption by the $A_{\rm V}$-derived $N_{\rm H}$. We note that more than one third of the X-ray spectra (31/78) in our sample are best fitted by model (1), while only one galaxy IC 4687 is better described by the two-component APEC model. Ideally, the data should be fitted with multiple APEC components characterized by discrete temperatures and differing absorption column densities. However, due to the poor quality of the X-ray data for most objects in the sample, this would lead to severe over-fitting. 

\section{Results}\label{sec:results}

In the following sections, we primarily focus on the diffuse thermal X-ray luminosity ({\lx}) of the hot gas in the 0.5$-$2 keV band. We assume that {\lx} is equal to the APEC component in our best-fitting models of the point source-excluded X-ray spectra. The corrected X-ray luminosities, SFRs, and merger stages for our targets are presented in Table~\ref{tab:table1} and \ref{tab:table3}.

\begin{figure*}
\centering 
\includegraphics[scale=0.98]{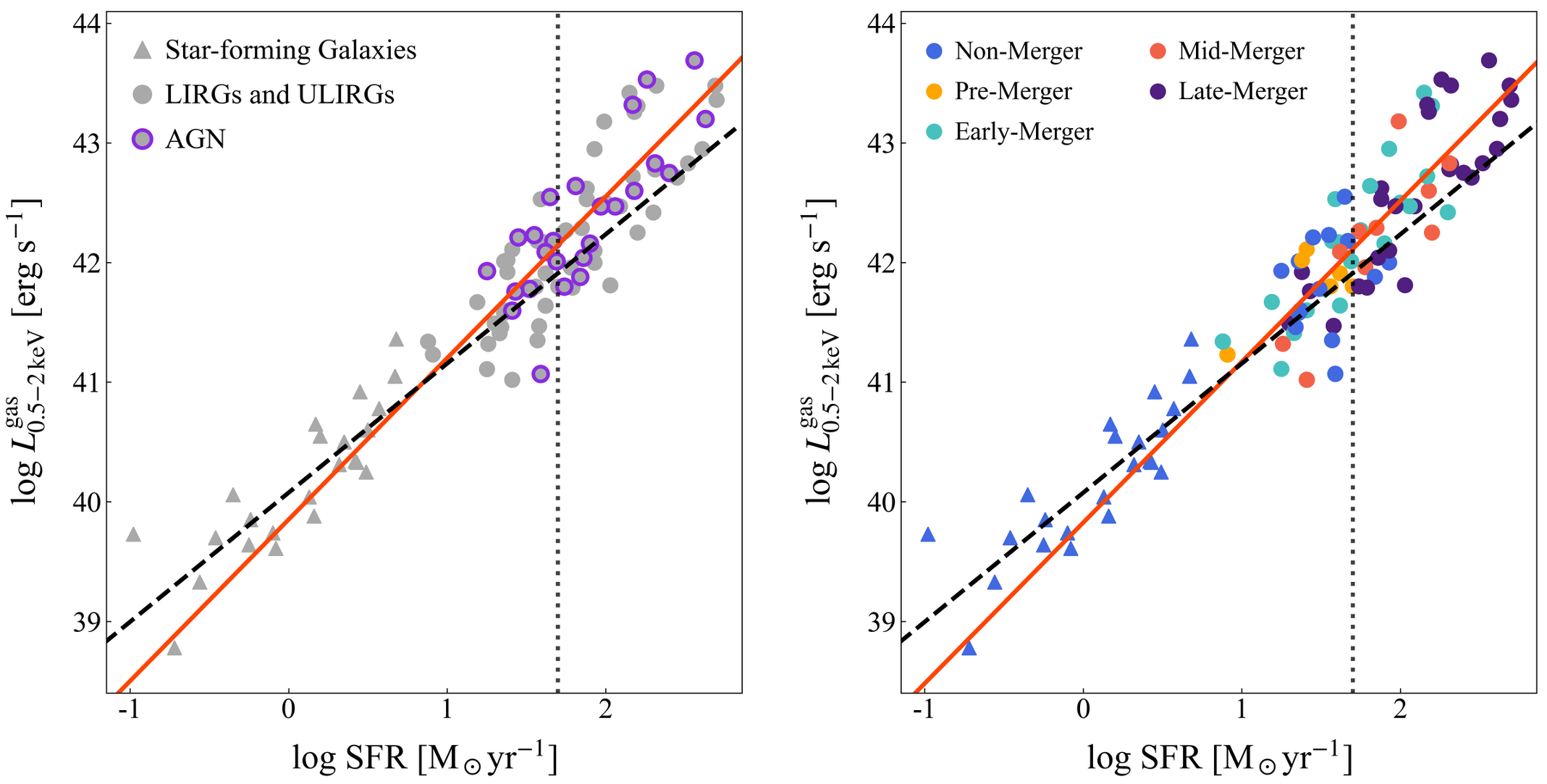}
\caption{Correlations between SFR and thermal X-ray luminosity of hot gas for star-forming galaxies, LIRGs, and ULIRGs. The X-ray luminosity has been corrected for intrinsic absorption in the 0.5$-$2 keV band. Left: objects hosting an AGN are marked with purple-bordered points. The vertical dotted line denotes SFR corresponding to 50 ${M_{\rm \odot}}\,\,{\rm yr^{-1}}$, which serves as a threshold for dividing the sample galaxies. The dashed line represents the linear relation for the star-forming galaxies (comparison sample) obtained by \citet{2025_Zhangcy_arXiv}, while the red solid line shows the best-fit relation for objects with SFR greater than 50 ${M_{\rm \odot}}\,\,{\rm yr^{-1}}$. Right: this panel presents the same data as the left one, but highlights the merger stages using multiple colors. The detailed classification of the stages can be found in Section~\ref{sec:Lx-SFR}. }
\vspace{0.55cm}
\label{fig:Lx-SFR-merger}
\end{figure*}

\subsection{Correlation between Hot Gas and SFR in LIRGs and ULIRGs}\label{sec:Lx-SFR}

The apparent luminosity of the diffuse emission from hot gas linearly correlates with SFR in both star-forming galaxies \citep{2003_Ranalli_A&A_39R, 2012_Mineo_hotgas, 2013_LiJiangTao_MNRAS_2} and (ultra)luminous infrared galaxies \citep{2005_Grimes_ApJ_187G, 2011_Iwasawa_A&A_106I}. However, most LIRGs and nearly all ULIRGs are merger remnants with intense, compact dust obscuration, resulting in strong absorption of the soft X-ray emission. In star-forming galaxies, only $\thicksim$10\% of the soft X-ray light is able to pass through the intrinsic column density \citep{2012_Mineo_hotgas, 2024_Zhangcy_ApJL}. Given the stronger intrinsic absorption in LIRGs and ULIRGs, it is essential to study the correlation between SFR and intrinsic thermal X-ray emission from these systems, as it helps in understanding star formation feedback in the intense star-forming environments. 

Figure~\ref{fig:Lx-SFR-merger} shows the {\lx}$-$SFR correlation with all X-ray luminosities corrected for the intrinsic absorption. We used 23 star-forming galaxies from \citet{2025_Zhangcy_arXiv} as a comparison sample. These galaxies are all nearby ($\textless$ 24 Mpc) and have no companions according to the NASA Extragalactic Database (NED)\footnote{\url{http://ned.ipac.caltech.edu}}. With the comparison sample, our measurements span $\thicksim$5 dex in X-ray luminosity (10$^{39}$$-$10$^{44}$ erg s$^{-1}$) and $\thicksim$4 dex in SFR (10$^{-1}$$-$10$^{3}$ ${M_{\rm \odot}}\,\,{\rm yr^{-1}}$). The SFR of our (U)LIRG sample is estimated using a combination of GALEX FUV and total infrared luminosity \citep[8-1000 $\mu$m;][]{2010_Howell_ApJ_572H}, with an uncertainty of typically less than ten percent.

The dashed line in Figure~\ref{fig:Lx-SFR-merger} is the linear {\lx}$-$SFR relation of the comparison sample fitted by \citet{2025_Zhangcy_arXiv} at the low star formation end (e.g., SFR $\textless$ 10 ${M_{\rm \odot}}\,\,{\rm yr^{-1}}$):
\begin{small}
\begin{equation}\label{eq1}
{\rm log}\,(\frac{L_{\rm 0.5-2\,keV}^{\rm gas}} {\rm erg\ s^{-1}}) = 1.08(\pm0.17) {\rm log}\,\frac{\rm SFR} {M_\odot\ {\rm yr}^{-1}} + 40.08(\pm0.08).  
\end{equation}
\end{small}
This relation is consistent with the results of \citet{2012_Mineo_hotgas} and \citet{2018_Smith_AJ_81S} after correction for the intrinsic absorption in their samples.

\begin{figure*}
\centering 
\includegraphics[scale=0.98]{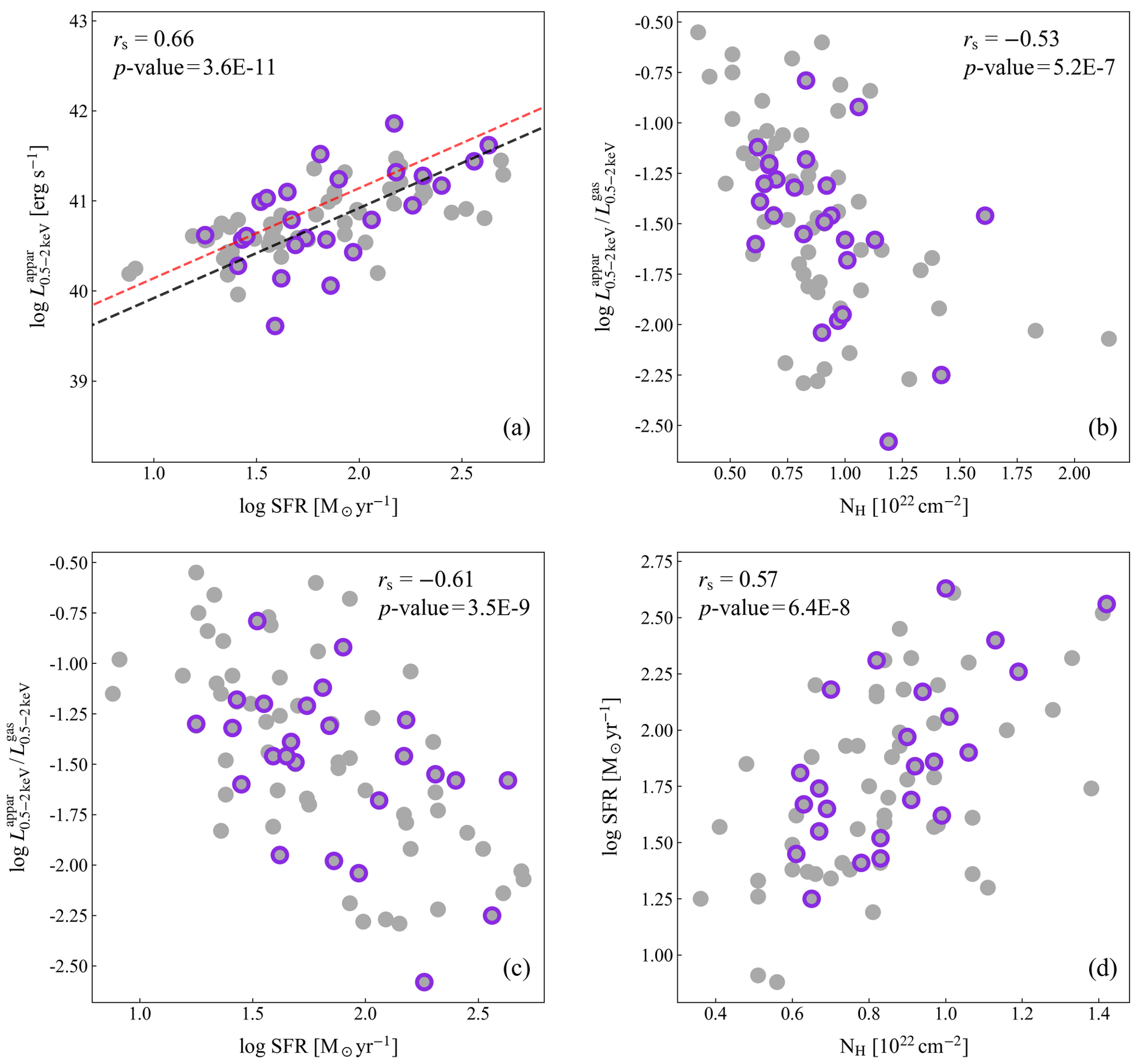}
\caption{Diffuse X-ray luminosity of hot gas versus intrinsic hydrogen column density and SFR. The black and red dashed lines in (a) represent the linear relations between the apparent hot gas luminosity and SFR for nearby star-forming galaxies \citep{2012_Mineo_hotgas} and the galactic coronae of edge-on galaxies \citep{2013_LiJiangTao_MNRAS_2}, respectively.           Point symbols are the same as those in Figure~\ref{fig:Lx-SFR-merger}. The Spearman rank correlation coefficient ($r_{\rm s}$) and the $p$-value are listed at the top of each panel. }
\vspace{0.5cm}
\label{fig:Lx-ratio-Av}
\end{figure*}

For SFR $\textless$ 50 ${M_{\rm \odot}}\,\,{\rm yr^{-1}}$ (the left of the dotted line in Figure~\ref{fig:Lx-SFR-merger}), the intrinsic {\lx} and SFR follow the linear relation well. However, at high SFRs, the thermal X-ray luminosity shows a steeper trend, with significant excess over the linear relation in the majority of our targets. Therefore, we re-quantified the {\lx}$-$SFR relation for objects with SFR greater than 50 ${M_{\rm \odot}}\,\,{\rm yr^{-1}}$ (high SFR, hereafter), using the Bayesian linear regression method as implemented in the LINMIX package \citep{2007_Kelly_ApJ_1489K}. The best-fit relation is 
\begin{small}
\begin{equation}\label{eq2}
{\rm log}\,(\frac{L_{\rm 0.5-2\,keV}^{\rm gas}} {\rm erg\ s^{-1}}) = 1.34(\pm0.23) {\rm log}\,\frac{\rm SFR} {M_\odot\ {\rm yr}^{-1}} + 39.82(\pm0.50).  
\end{equation}
\end{small}
The $rms$ scatter of this relation is $\sigma$ = 0.38 dex, with a Spearman rank correlation coefficient of $r_{\rm s}$ = 0.72. The right panel of Figure~\ref{fig:Lx-SFR-merger} shows the same figure but with the targets color-coded by their merger stages. Information on the merger stage was obtained from \citet{2013_Stierwalt_ApJS_1S}, based on visual inspection of IRAC 3.6 $\mu$m images. The stage classifications are: (1) non-merger, (2) pre-merger (galaxy pairs before first encounter), (3) early-merger (after first encounter, with symmetric and intact disks but visible tidal tails), (4) mid-merger (featuring disrupted or amorphous disks, tidal tails, and other signs of interaction), and (5) late-merger (two nuclei embedded in a common envelope). More than half of the objects with high SFR are in late-merger, followed by early- and mid-merger. This suggests that the large scatter of the {\lx}$-$SFR relation in the (U)LIRGs can be attributed to the merger activities, which typically involve long timescales (greater than 1 Gyr) and strongly affect the evolution of hot gas \citep{2018_Smith_AJ_81S}.

In Figure~\ref{fig:Lx-SFR-merger}, we use purple circles to highlight the objects hosting AGN (see Table~\ref{tab:table1}, and also see \citealt{2011_Iwasawa_A&A_106I} and \citealt{2018_Torres_A&A_140T} for the detailed source classification) and find no significant excess in thermal X-ray luminosity compared to the non-AGN galaxies. Moreover, we note that AGN exist throughout all stages of the merger process within the sample. This indicates that the heating from AGN does not govern the evolution of the hot gas on a galactic scale, which is instead dominated by star formation processes (e.g., supernovae and stellar winds). Similar results have been reported in a systematic study of 49 major mergers \citep{2018_Smith_AJ_81S}, as well as in the Seyfert 2 galaxy M51 \citep{2025_Zhangcy_ApJ_15Z}. In their work, the influence of AGN on the intrinsic X-ray luminosity of hot gas is not significant on either resolved or global scales.

The slope of the {\lx}$-$SFR relation for the high SFR targets is consistent with the super-linear relation observed in the central regions ($\thicksim$0.3$-$2 kpc) of the comparison sample, as reported by \citet{2025_Zhangcy_arXiv}. This suggests that the global environments of objects with high SFR have similar characteristics to the central regions of normal spiral galaxies. It can be inferred that the merger of galaxies provides substantial amounts of gas, triggering bursts of star formation not only in the nuclear region of the system but also throughout the galactic disk. Subsequently, as the system reaches a high level of star formation (e.g., global SFR $\textgreater$ 50 ${M_{\rm \odot}}\,\,{\rm yr^{-1}}$), large amounts of gas distributed across the galactic disk are intensely heated, and this eventually leads to the excess of the intrinsic X-ray luminosity. To further understand the origin and heating mechanisms of hot gas, observations with higher sensitivity and spatial resolution are essential, but such observations are beyond the scope of this work. 

\subsection{Hot Gas Luminosities for Galactic and Intrinsic Absorption}\label{sec:Gal_Int}

In this section, we first examine the correlation between SFR and the thermal X-ray luminosity corrected only for Galactic absorption. Figure~\ref{fig:Lx-ratio-Av}(a) shows that, compared to the slightly higher relation for galactic coronae reported by \citet{2013_LiJiangTao_MNRAS_2}, the objects are consistent with the scaling relation of $L_{\rm X}$/SFR $\thicksim$ 8.3 $\cdot$ 10$^{38}$ (erg s$^{-1}$)/(${M_{\rm \odot}}\,\,{\rm yr^{-1}}$) established by \citet{2012_Mineo_hotgas}. This could be attributed to the fact that the diffuse emission of our sample galaxies is mainly from the galactic disk. However, the significant dispersion in these data points suggests that both the origin and absorption of hot gas emission in (U)LIRGs involve complex and diverse physical processes. In addition, the inclination of galaxies is likely to play a role in this dispersion, but a detailed investigation lies beyond the scope of this study because most of the galaxies in our sample are mergers and are located at relatively large distances.

We subsequently analyzed the ratio between the apparent luminosity ($L_{\rm 0.5 - 2\,keV}^{\rm appar}$) and the intrinsic luminosity ({\lx}). Figures~\ref{fig:Lx-ratio-Av}(b) and (c) show this ratio as a function of intrinsic $N_{\rm H}$ and SFR, respectively. As expected, the $L_{\rm 0.5 - 2\,keV}^{\rm appar}$/{\lx} ratio declines with increasing $N_{\rm H}$, reflecting stronger absorption at higher column densities. Notably, a similar decreasing trend is observed with SFR, although the $L_{\rm 0.5 - 2\,keV}^{\rm appar}$/{\lx}$-$SFR relation shows considerable dispersion. A Spearman rank test analysis yields a correlation coefficient of $r_{\rm s}$ = $-$0.53 ($p$-value = 5.2$\times$10$^{-7}$) for $N_{\rm H}$, and $r_{\rm s}$ = $-$0.61 ($p$-value = 3.5$\times$10$^{-9}$) for SFR, indicating statistically significant negative correlations in both cases.

We next explored the correlation between the intrinsic $N_{\rm H}$ and SFR. The column density includes both atomic and molecular hydrogen (e.g., $N_{\rm H}$ = $N_{\rm H_{I}}$ + 2$N_{\rm H_{2}}$), where the molecular hydrogen is the most abundant molecule in the Universe and the primary raw material for star formation \citep{2008_Bigiel_AJ_2846B,2013_Schruba_IAUS_311S}. Figure~\ref{fig:Lx-ratio-Av}(d) indeed shows a moderate positive correlation between $N_{\rm H}$ and SFR, with $r_{\rm s}$ = 0.57 and $p$-value = 6.4$\times$10$^{-8}$. The large dispersion in the SFR$-$$N_{\rm H}$ correlation could be attributed to variations in the fraction of atomic gas within the total column density.

Comparing the $L_{\rm 0.5 - 2\,keV}^{\rm appar}$/{\lx}$-$$N_{\rm H}$ relation, we infer that the correlation between $L_{\rm 0.5 - 2\,keV}^{\rm appar}$/{\lx} and SFR is primarily driven by or related to the correlation of $N_{\rm H}$ with SFR, since the soft X-ray absorption is directly affected by the column density.

\subsection{Contribution of Unresolved Compact Sources}\label{sec:pointsources}

\begin{figure}
\includegraphics[scale=1]{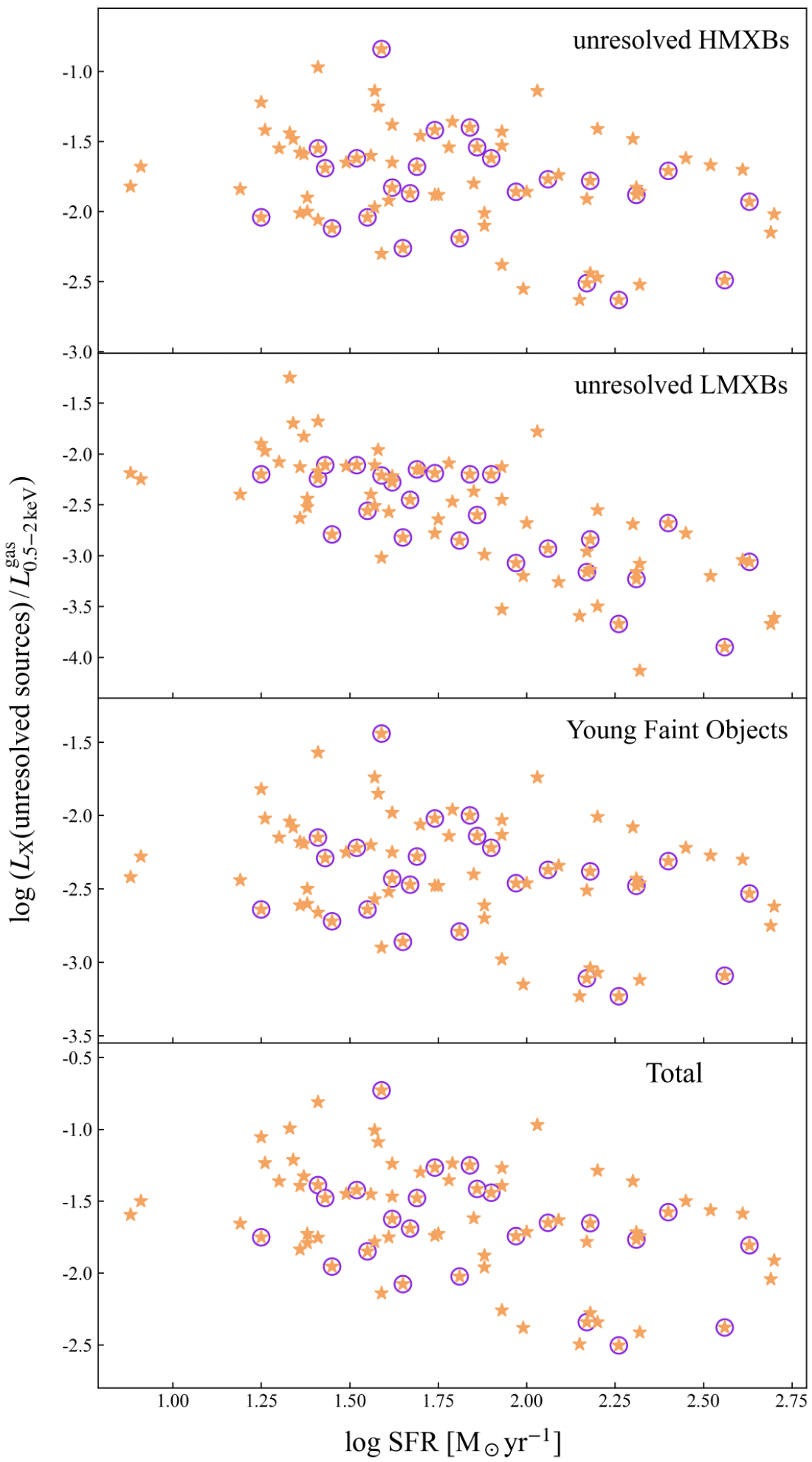}
\caption{Predicted 0.5$-$2 keV luminosities of unresolved HMXBs, LMXBs, young faint objects, as well as their total, normalized by the intrinsic thermal luminosity {\lx} and plotted against the SFR. Galaxies hosting an AGN are marked with purple circles. }
\vspace{0.15cm}
\label{fig:unresolved-sources}
\end{figure}

For the observed diffuse X-ray spectra, several intrinsically faint compact sources below the point source detection threshold could contribute a non-thermal component. The unresolved sources include high-mass X-ray binaries (HMXBs), low-mass X-ray binaries (LMXBs), cataclysmic variables (CVs), coronally active binaries (ABs), and young faint objects, including protostars and young stars. For unresolved young supernova remnants, they are expected to have a thermal X-ray spectrum \citep{2010_Long_ApJS_495L} and thus are incorporated within the hot gas component of our analysis.

The contribution from unresolved HMXBs is scaled with SFR \citep{2003_Grimm_MNRAS_793G, 2012_Mineo_HMXBs}. Thus, we estimated the total luminosities of these sources following the method of \citet{2018_Smith_AJ_81S}:
\begin{small}
\begin{equation}\label{eq3}
    L_{\mathrm{X}}(\mathrm{unresolved\,HMXBs})=3.725\,\mathrm{SFR}[L^{0.4}_{\mathrm{upper}}-0.025]. 
\end{equation}
\end{small}
where a power-law X-ray spectrum is assumed and $L_{\rm upper}$ is the luminosity of HMXBs in units of 10$^{38}$ erg s$^{-1}$. To obtain the upper limit of the unresolved luminosities, we set $L_{\rm upper}$ equal to 10 $\times$ 10$^{38}$ erg s$^{-1}$, with assuming a photon index of $\Gamma$ = 1.8, which is typical for point sources in nearby galaxies \citep{2002_Kong_ApJ_738K, 2004_Swartz_ApJS_519S}. 

The emission from unresolved LMXBs is expected to scale with the stellar mass \citep{2004_Gilfanov_MNRAS_146G, 2011_Boroson_ApJ_12B}. We calculated the X-ray luminosity from unresolved LMXBs using the prescription of \citet{2004_Gilfanov_MNRAS_146G}:
\begin{small}
\begin{equation}\label{eq4}
    L_{\mathrm{X}}(\mathrm{unresolved\,LMXBs})=(8 \pm 0.5) \times 10^{39}\,\rm{erg\,s^{-1}\,per}\,10^{11} M_{\rm \odot}. 
\end{equation}
\end{small}

For young faint objects, \citet{2011_Bogdan_MNRAS_1901B} found that the hard X-ray luminosity in the 2$-$10 keV band of these objects shows a tight correlation with SFR:
\begin{small}
\begin{equation}\label{eq5}
    L_{\mathrm{X}}(\mathrm{young})/{\rm SFR}=(1.7 \pm 0.9) \times 10^{38} (\mathrm{erg \, s^{-1}})/({\rm M_{\odot}\ yr^{-1}}). 
\end{equation}
\end{small}
Following \citet{2025_Kyritsis_A&A_128K}, we adopted a conversion factor of 0.64 to estimate the emission in the 0.5$-$2 keV band from the hard X-ray luminosities.

\citet{2008_Revnivtsev_A&A_37R} found that the total X-ray luminosity of ABs and CVs follows a constant ratio of log($L_{\rm X}/L_{\rm K}$) $\thicksim$ $-$5.8 relative to the $K$-band emission of galaxies, indicating a minor contribution to the global X-ray emission. \citet{2011_Boroson_ApJ_12B} also reported that, on average, the X-ray luminosity of LMXBs is approximately ten times greater than that of ABs and CVs. Therefore, we conclude that the contribution of ABs and CVs to the observed diffuse X-ray emission in the sample galaxies is negligible.

In Figure~\ref{fig:unresolved-sources}, we plot the SFR against the luminosities of these unresolved sources, individually and in total, normalized by the intrinsic thermal X-ray luminosity {\lx}. We find that the normalized luminosities of each type of source show a decreasing trend with SFR. Compared to the diffuse luminosity in the 0.5$-$2 keV band, the total contribution of unresolved sources exceeds 10\% in only 3 out of 78 objects, with the highest reaching 19\%. These small contributions have little impact on our results and thus can be safely ignored. In conclusion, the luminosity derived from the point source-excluded X-ray spectra is a reasonable estimate of the hot gas emission.

\section{Summary}\label{sec:summary}

In this paper, we study the properties of hot ionized interstellar gas in 78 nearby (U)LIRGs selected from the C-GOALS program. Nearly all galaxies in the sample (64/78) are undergoing mergers at various stages of interaction. For the extreme star-forming environments, surrounded by a substantial amount of gaseous material, we use a dust extinction method to estimate the intrinsic $N_{\rm H}$ for these systems. After correcting for the intrinsic absorption, the diffuse hot gas luminosities of most objects with high SFRs ($\textgreater$ 50 ${M_{\rm \odot}}\,\,{\rm yr^{-1}}$) show a significant excess over the predictions of the standard linear $L_{\rm X}$$-$SFR relation. The Bayesian method gives a super-linear relation between {\lx} and SFR for the high-SFR objects with a slope of 1.34. However, the scatter of this relation is relatively large ($\thicksim$ 0.38 dex) compared to that observed in normal star-forming galaxies, possibly due to the strong influence of mergers. The heating from AGN appears to have little effect on the global hot gas luminosities throughout all five merger stages. Thus, this super-linear relation suggests a scenario in which the merger of galaxies delivers substantial gas, triggering widespread star formation in both the nuclear region and the galactic disk, ultimately enhancing the global hot gas emission and leading to the excess of X-ray luminosity. 

The ratio of $L_{\rm 0.5 - 2\,keV}^{\rm appar}$ to {\lx} shows statistically significant negative correlations with intrinsic $N_{\rm H}$ and SFR. However, in contrast to the luminosity ratio, the SFR shows a moderate positive correlation with the $N_{\rm H}$ ($r_{\rm s}$ = 0.57). Thus, compared to the $L_{\rm 0.5 - 2\,keV}^{\rm appar}$/{\lx}$-$$N_{\rm H}$ relation, the correlation between $L_{\rm 0.5 - 2\,keV}^{\rm appar}$/{\lx} and SFR may be driven by the underlying SFR$-$$N_{\rm H}$ relation, since the observed X-ray luminosity depends strongly on the absorbing column density. 


We thank the anonymous referee for providing helpful comments that improved our work. We acknowledge support from the National Key R$\&$D Program of China (grant No. 2023YFA1607904) and the National Natural Science Foundation of China (NSFC) grants 12033004, 12221003, and 12333002 and the science research grant from CMS-CSST-2025-A10 and CMS-CSST-2025-A07. This paper employs a list of \emph{Chandra} datasets, obtained by the \emph{Chandra} X-ray Observatory, contained in the \emph{Chandra} Data Collection (CDC) 439\dataset[doi:10.25574/cdc.439]{https://doi.org/10.25574/cdc.439}. This research has also made use of the NASA/IPAC Extragalactic Database (NED), which is operated by the Jet Propulsion Laboratory, California Institute of Technology, under contract with the National Aeronautics and Space Administration. The Chandra data archive is maintained by the Chandra X-ray Center at the Smithsonian Astrophysical Observatory. CARTA is the Cube Analysis and Rendering Tool for Astronomy, a new image visualization and analysis tool designed for the ALMA, the VLA, and the SKA pathfinders. This work makes use of data from GOALS team. GOALS is combining imaging and spectroscopic data from NASA's Spitzer, Hubble, Chandra and GALEX space-borne observatories in a comprehensive study of over 200 of the most luminous infrared-selected galaxies in the local Universe. 


\bibliography{bibliography}{}
\bibliographystyle{aasjournal}

\appendix
\startlongtable
\begin{deluxetable*}{lcccccccccc}
\renewcommand{\arraystretch}{0.99}  
\vspace{-0.9cm}
\centering
\tablecaption{Galaxy Sample \label{tab:table1}}
\addtolength{\tabcolsep}{1.5pt}
\tablehead{
\colhead{Galaxy} & \colhead{R.A. (J2000)} & \colhead{Decl. (J2000)} & \colhead{$z$} & \colhead{$D_L$} & \colhead{IRX} & \colhead{SFR} & \colhead{Stage} & \colhead{$N_{\rm H,Gal}$} & \colhead{AGN} \\
\colhead{(1)} & \colhead{(2)} & \colhead{(3)} & \colhead{(4)} & \colhead{(5)} & \colhead{(6)} & \colhead{(7)} & \colhead{(8)} & \colhead{(9)} & \colhead{(10)} 
}
\startdata
IRAS F01364--1042 & 01h38m52.92s & --10d27m11.4s & 0.0483 & 210.0 & 3.136 & 122.61 & d & 2.0 & ... \\
ESO 255-IG7 & 06h27m22.45s & --47d10m48.7s & 0.0388 & 173.0 & 2.048 & 140.00 & b & 3.8 & ... \\
ESO 60-IG16 & 08h52m31.29s & --69d01m57.0s & 0.0463 & 210.0 & 2.495 & 115.40 & b & 5.2 & Y \\
IRAS 09022--3615 & 09h04m12.70s & --36d27m01.1s & 0.0596 & 271.0 & 3.451 & 359.05 & d & 26.6 & Y \\
IRAS F09111--1007 & 09h13m37.61s & --10d19m24.8s & 0.0541 & 246.0 & 2.619 & 198.46 & b & 4.6 & ... \\
ESO 374-IG32 & 10h06m04.80s & --33d53m15.0s & 0.0341 & 156.0 & 2.394 & 106.13 & d & 8.8 & ... \\
VV 340 N & 14h57m00.70s & +24d37m05.8s & 0.0337 & 157.0 & 2.610 & 79.04 & b & 3.3 & Y \\
IRAS F17132+5313 & 17h14m20.00s & +53d10m30.0s & 0.0509 & 232.0 & 2.435 & 159.67 & b & 1.9 & ... \\
ESO 593-IG8 & 19h14m30.90s & --21d19m07.0s & 0.0487 & 222.0 & 2.209 & 150.38 & d & 8.1 & ... \\
IRAS F19297--0406 & 19h32m21.25s & --03d59m56.3s & 0.0857 & 395.0 & 4.401 & 494.79 & d & 15.1 & ... \\
CGCG 448-020 & 20h57m23.90s & +17d07m39.0s & 0.0361 & 161.0 & 1.662 & 156.77 & c & 6.9 & ... \\
IRAS F22491--1808 & 22h51m49.26s & --17d52m23.5s & 0.0778 & 351.0 & 2.198 & 279.16 & d & 2.3 & ... \\
ESO 77-IG14 & 23h21m04.53s & --69d12m54.2s & 0.0416 & 186.0 & 2.856 & 100.48 & b & 3.3 & ... \\
IRAS F05189-2524 & 05h21m01.47s & --25d21m45.4s & 0.0425 & 187.0 & 2.783 & 253.48 & d & 1.7 & Y \\
UGC 4881 & 09h15m55.11s & +44d19m54.1s & 0.0393 & 178.0 & 2.197 & 97.13 & c & 1.4 & ... \\
UGC 5101 & 09h35m51.65s & +61d21m11.3s & 0.0394 & 177.0 & 2.925 & 180.18 & d & 3.0 & Y \\
IRAS F10565+2448 & 10h59m18.14s & +24d32m34.3s & 0.0431 & 197.0 & 3.247 & 209.13 & d & 1.1 & ... \\
NGC 3690 & 11h28m32.25s & +58d33m44.0s & 0.0104 & 50.7 & 1.771 & 150.55 & c & 0.9 & Y \\
IRAS F12112+0305 & 12h13m46.00s & +02d48m38.0s & 0.0733 & 340.0 & 2.510 & 402.89 & d & 1.8 & ... \\
IRAS F14348-1447 & 14h37m38.37s & --15d00m22.8s & 0.0827 & 387.0 & 2.466 & 428.54 & d & 7.5 & Y \\
VV 705 & 15h18m06.28s & +42d44m41.2s & 0.0402 & 183.0 & 2.061 & 147.81 & b & 1.8 & ... \\
IRAS F15250+3608 & 15h26m59.40s & +35d58m37.5s & 0.0552 & 254.0 & 2.264 & 211.11 & d & 1.5 & ... \\
UGC 9913 & 15h34m57.12s & +23d30m11.5s & 0.0182 & 87.9 & 3.423 & 327.74 & d & 3.9 & ... \\
NGC 6240 & 16h52m58.89s & +02d24m03.4s & 0.0245 & 116.0 & 2.336 & 148.44 & d & 4.9 & Y \\
IRAS F17207-0014 & 17h23m21.96s & --00d17m00.9s & 0.0428 & 198.0 & 5.154 & 501.22 & d & 9.7 & ... \\
ESO 286-IG19 & 20h58m26.79s & --42d39m00.3s & 0.0430 & 193.0 & 2.108 & 203.56 & d & 3.3 & ... \\
ESO 148-IG2 & 23h15m46.78s & --59d03m15.6s & 0.0446 & 199.0 & 2.054 & 204.60 & c & 1.6 & Y \\
UGC 08387 & 13h20m35.34s & +34d08m22.2s & 0.0233 & 110.0 & 2.379 & 94.18 & d & 1.0 & Y \\
CGCG 436--030 & 01h20m02.72s & +14d21m42.9s & 0.0312 & 134.0 & 1.974 & 85.87 & b & 3.4 & ... \\
NGC 0695 & 01h51m14.24s & +22d34m56.5s & 0.0325 & 139.0 & 2.053 & 84.64 & N & 6.9 & ... \\
CGCG 043--099 & 13h01m50.80s & +04d20m00.0s & 0.0375 & 175.0 & 2.330 & 84.73 & d & 1.9 & ... \\
MCG+07--23--019 & 11h03m53.20s & +40d50m57.0s & 0.0345 & 158.0 & 1.761 & 75.13 & d & 1.0 & ... \\
NGC 6670 (E) & 18h33m37.67s & +59d53m21.3s & 0.0289 & 129.5 & 2.229 & 41.39 & b & 3.9 & ... \\
NGC 6670 (W) & 18h33m32.78s & +59d53m11.7s & 0.0289 & 129.5 & 2.558 & 36.78 & b & 3.9 & ... \\
NGC 5331 (N) & 13h52m16.43s & +02d06m30.9s & 0.033 & 155.0 & 1.398 & 18.10 & c & 2.0 & ... \\
NGC 5331 (S) & 13h52m16.21s & +02d06m05.1s & 0.033 & 155.0 & 2.393 & 60.78 & c & 2.0 & ... \\
IC 2810 (NW) & 11h25m45.07s & +14d40m36.0s & 0.0342 & 157.0 & 2.266 & 49.68 & a & 2.5 & ... \\
NGC 3256 & 10h27m51.27s & --43d54m13.8s & 0.0094 & 38.9 & 2.299 & 76.46 & d & 9.1 & ... \\
IRAS F16164--0746 & 16h19m11.79s & --07d54m02.8s & 0.0272 & 128.0 & 2.573 & 72.75 & d & 11.3 & Y \\
IC 4687 & 18h13m39.80s & --57d43m30.7s & 0.0173 & 81.9 & 2.241 & 38.51 & b & 11.5 & ... \\
IC 4689 & 18h13m40.38s & --57d44m54.3s & 0.0173 & 81.9 & 2.176 & 15.49 & b & 11.5 & ... \\
IC 5298 & 23h16m00.70s & +25d33m24.1s & 0.0274 & 119.0 & 2.451 & 69.67 & N & 5.7 & Y \\
NGC 6090 & 16h11m40.70s & +52d27m24.0s & 0.0293 & 137.0 & 1.325 & 71.10 & c & 1.6 & ... \\
NGC 5256 & 13h38m17.52s & +48d16m36.7s & 0.0279 & 129.0 & 1.679 & 65.00 & b & 1.7 & Y \\
IRAS F03359+1523 & 03h38m46.70s & +15d32m55.0s & 0.0354 & 152.0 & 2.562 & 61.22 & d & 13.8 & ... \\
ESO 550--IG025 & 04h21m20.02s & --18d48m47.6s & 0.0322 & 135.8 & 2.142 & 56.46 & b & 3.2 & ... \\
NGC 0034 & 00h11m06.55s & +12d06m26.3s & 0.0196 & 84.1 & 1.799 & 55.25 & d & 2.1 & Y \\
MCG+12--02--001 & 00h54m03.61s & +73d05m11.8s & 0.0157 & 69.8 & 3.593 & 54.49 & c & 22.0 & ... \\
ESO 440--IG058 (S) & 12h06m51.87s & --31d56m59.2s & 0.0232 & 112.0 & 2.819 & 41.15 & b & 5.6 & ... \\
NGC 7130 & 21h48m19.50s & --34d57m04.7s & 0.0162 & 72.7 & 1.712 & 46.88 & N & 1.9 & Y \\
NGC 7771 & 23h51m24.80s & +20d06m42.2s & 0.0143 & 61.2 & 2.214 & 25.79 & a & 4.0 & ... \\
NGC 7770 & 23h51m22.55s & +20d05m49.2s & 0.0143 & 61.2 & 1.396 & 8.16 & a & 4.0 & ... \\
NGC 7592 (E) & 23h18m22.60s & --04d24m58.0s & 0.0244 & 106.0 & 0.991 & 17.96 & b & 3.8 & ... \\
NGC 7592 (W) & 23h18m21.78s & --04d24m57.0s & 0.0244 & 106.0 & 2.097 & 25.85 & b & 3.8 & Y \\
NGC 6285 & 16h58m23.99s & +58d57m21.7s & 0.0183 & 85.7 & 1.532 & 7.58 & b & 1.8 & ... \\
NGC 6286 & 16h58m31.63s & +58d56m13.3s & 0.0183 & 85.7 & 2.204 & 33.01 & b & 1.8 & Y \\
NGC 4922 (N) & 13h01m25.27s & +29d18m49.5s & 0.0236 & 111.0 & 2.615 & 41.61 & c & 0.9 & Y \\
NGC 3110 & 10h04m02.11s & --06d28m29.2s & 0.0169 & 79.5 & 1.646 & 41.72 & a & 3.5 & ... \\
NGC 0232 & 00h42m45.82s & --23d33m40.9s & 0.0227 & 95.2 & 2.404 & 48.65 & b & 1.4 & Y \\
MCG+08--18--013 (E) & 09h36m37.19s & +48d28m27.7s & 0.0259 & 117.0 & 2.066 & 36.24 & a & 1.7 & ... \\
CGCG 049--057 & 15h13m13.09s & +07d13m31.8s & 0.013 & 65.4 & 4.156 & 39.16 & N & 2.6 & Y \\
NGC 1068 & 02h42m40.71s & --00d00m47.8s & 0.0038 & 15.9 & 1.865 & 44.29 & N & 2.9 & Y \\
UGC 02238 & 02h46m17.49s & +13d05m44.4s & 0.0219 & 92.4 & 2.591 & 37.70 & d & 8.9 & ... \\
MCG--03--34--064 & 13h22m24.46s & --16d43m42.9s & 0.0165 & 82.2 & 2.204 & 27.08 & d & 5.0 & Y \\
ESO 350--IG038 & 00h36m52.25s & --33d33m18.1s & 0.0206 & 89.0 & 1.128 & 37.02 & N & 2.4 & ... \\
MCG--01--60--022 & 23h42m00.85s & --03d36m54.6s & 0.0232 & 100.0 & 1.654 & 28.44 & N & 3.6 & Y \\
NGC 5135 & 13h25m44.06s & --29d50m01.2s & 0.0137 & 60.9 & 1.804 & 35.41 & N & 4.9 & Y \\
IC 5179 & 22h16m09.10s & --36d50m37.4s & 0.0114 & 51.4 & 1.619 & 31.01 & N & 1.4 & ... \\
CGCG 465--012 & 03h54m16.08s & +15d55m43.4s & 0.0222 & 94.3 & 2.920 & 20.18 & d & 14.8 & ... \\
MCG--02--33--098--9 & 13h02m19.70s & --15d46m03.0s & 0.0159 & 78.7 & 1.973 & 25.87 & c & 3.7 & ... \\
ESO 343--IG013 & 21h36m10.83s & --38d32m37.9s & 0.0191 & 85.8 & 2.014 & 24.22 & d & 2.8 & ... \\
NGC 2146 & 06h18m37.71s & +78d21m25.3s & 0.003 & 17.5 & 2.820 & 22.87 & N & 7.1 & ... \\
NGC 5653 & 14h30m10.42s & +31d12m55.8s & 0.0119 & 60.2 & 1.640 & 24.24 & a & 1.3 & ... \\
NGC 0023 & 00h09m53.41s & +25d55m25.6s & 0.0152 & 65.2 & 1.724 & 23.34 & N & 3.4 & ... \\
NGC 7552 & 23h16m10.77s & --42d35m05.4s & 0.0054 & 23.5 & 1.793 & 22.98 & N & 1.2 & ... \\
NGC 1961 & 05h42m04.65s & +69d22m42.4s & 0.0131 & 59.0 & 1.403 & 21.38 & b & 8.1 & ... \\
NGC 1365 & 03h33m36.37s & --36d08m25.4s & 0.0055 & 17.9 & 1.760 & 17.83 & N & 1.3 & Y \\
NGC 3221 & 10h22m19.98s & +21d34m10.5s & 0.0137 & 65.7 & 1.896 & 21.75 & N & 1.9 & ... \\
\enddata
\tablecomments{Column (2): right ascension (J2000) taken from NED as of October 2008. Column (3): declination (J2000) taken from NED as of October 2008. Column (4): the best available heliocentric redshift in NED as of October 2008. Column (5): luminosity distance, in units of Mpc \citep{2009_Armus_PASP_559A}. Column (6): total infrared to FUV luminosity ratio \citep{2010_Howell_ApJ_572H}. Column (7): star formation rate, in units of $M_{\odot}$ ${\rm yr}^{-1}$ \citep{2010_Howell_ApJ_572H}. Column (8): merger stage \citep{2013_Stierwalt_ApJS_1S}. Column (9): Galactic absorption column density, in units of 10$^{20}$cm$^{-2}$. Column (10): Y indicates a source hosting an AGN. }
\end{deluxetable*}

\begin{deluxetable*}{lcclcc}
\centering
\tablecaption{$Chandra$ Data \label{tab:table2}}
\addtolength{\tabcolsep}{3.0pt}
\tablehead{
\colhead{Galaxy} & \colhead{Exp. Time} & \colhead{Obs. ID} & \colhead{Galaxy} & \colhead{Exp. Times} & \colhead{Obs. ID} \\
\colhead{} & \colhead{(ks)} & \colhead{} & \colhead{} & \colhead{(ks)} & \colhead{} 
}
\startdata
IRAS F01364--1042 & 14.57 & 7801 & IC 4687 & 14.48 & 15056 \\
ESO 255-IG7 & 14.57 & 7803 & IC 4689 & 14.48 & 15056 \\
ESO 60-IG16 & 14.57 & 7888 & IC 5298 & 14.78 & 15059 \\
IRAS 09022--3615 & 14.85 & 7805 & NGC 6090 & 14.79 & 6859 \\
IRAS F09111--1007 & 14.63 & 7806 & NGC 5256 & 19.69 & 2044 \\
ESO 374-IG32 & 14.36 & 7807 & IRAS F03359+1523 & 14.76 & 6856 \\
VV 340 N & 14.57 & 7812 & ESO 550--IG025 & 14.78 & 15060 \\
IRAS F17132+5313 & 14.85 & 7814 & NGC 0034 & 14.78 & 15061 \\
ESO 593-IG8 & 14.97 & 7816 & MCG+12--02--001 & 14.31 & 15062 \\
IRAS F19297--0406 & 16.42 & 7890 & ESO 440--IG058 (S) & 14.78 & 15064 \\
CGCG 448-020 & 14.57 & 7818 & NGC 7130 & 38.64 & 2188 \\
IRAS F22491--1808 & 14.97 & 7821 & NGC 7771 & 16.71 & 10397 \\
ESO 77-IG14 & 14.98 & 7822 & NGC 7770 & 16.71 & 10397 \\
IRAS F05189-2524 & 34.55 & 2034, 3432 & NGC 7592 (E) & 14.99 & 6860 \\
UGC 4881 & 14.77 & 6857 & NGC 7592 (W) & 14.99 & 6860 \\
UGC 5101 & 49.32 & 2033 & NGC 6285 & 14.00 & 10566 \\
IRAS F10565+2448 & 28.87 & 3952 & NGC 6286 & 14.00 & 10566 \\
NGC 3690 & 124.86 & 1641, 6227, & NGC 4922 (N) & 24.42 & 4775, 15065, \\
 & & 15077, 15619 & & & 18201 \\
IRAS F12112+0305 & 28.96 & 4110, 4934 & NGC 3110 & 14.87 & 15069 \\
IRAS F14348--1447 & 14.72 & 6861 & NGC 0232 & 21.65 & 12872, 15066 \\
VV 705 & 14.47 & 6858 & MCG+08--18--013 (E) & 13.79 & 15067 \\
IRAS F15250+3608 & 9.84 & 4112 & CGCG 049--057 & 19.06 & 10399 \\
UGC 9913 & 292.42 & 869, 1609(2-3) & NGC 1068 & 47.44 & 344 \\
NGC 6240 & 182.05 & 1590, 12713 & UGC 02238 & 14.87 & 15068 \\
IRAS F17207--0014 & 57.72 & 2035, 4114 & MCG--03--34--064 & 71.49 & 7373, 23690, 25253, \\
 & & & & & 27267, 2780(2--3) \\
ESO 286-IG19 & 44.87 & 2036 & ESO 350--IG038 & 127.23 & 8175, 1669(5--7) \\
ESO 148-IG2 & 49.31 & 2037 & MCG--01--60--022 & 18.90 & 10570 \\
UGC 08387 & 14.07 & 7811 & NGC 5135 & 29.30 & 2187 \\
CGCG 436--030 & 13.82 & 15047 & IC 5179 & 11.96 & 10392 \\
NGC 0695 & 14.78 & 15046 & CGCG 465--012 & 14.87 & 15071 \\
CGCG 043--099 & 14.78 & 15048 & MCG--02--33--098--9 & 14.87 & 15072 \\
MCG+07--23--019 & 52.34 & 12977 & ESO 343--IG013 & 14.78 & 15073 \\
NGC 6670 (E) & 14.77 & 15049 & NGC 2146 & 63.08 & 313(1--6), 24706 \\
NGC 6670 (W) & 14.77 & 15049 & NGC 5653 & 16.52 & 10396 \\
NGC 5331 (N) & 14.78 & 15051 & NGC 0023 & 19.45 & 10401 \\
NGC 5331 (S) & 14.78 & 15051 & NGC 7552 & 200.94 & 7848, 2026(7--8), \\
& & & & & 2167(5--6) \\
IC 2810 (NW) & 14.78 & 15053 & NGC 1961 & 32.83 & 10531 \\
NGC 3256 & 70.37 & 835, 3569, 16026 & NGC 1365 & 101.93 & 3554, 68(68--73) \\
IRAS F16164--0746 & 14.78 & 15057 & NGC 3221 & 28.98 & 10398, 19334 \\
\enddata
\tablecomments{The exposure times are good time intervals. }
\end{deluxetable*}

\startlongtable
\begin{deluxetable*}{lcccccc}
\renewcommand{\arraystretch}{0.99}  
\centering
\tablecaption{Hot Gas X-ray Spectral Properties \label{tab:table3}}
\addtolength{\tabcolsep}{5.0pt}
\tablehead{
\colhead{Galaxy} & \colhead{$N_{\rm H}$} & \colhead{Model} & \colhead{$L_{\rm 0.5 - 2\,keV}^{\rm appar}$} & \colhead{\lx} & \colhead{$kT$} & \colhead{C-stat/d.o.f} \\
\colhead{} & \colhead{(10$^{22}$cm$^{-2}$)} & \colhead{} & \colhead{(erg s$^{\rm -1}$)} & \colhead{(erg s$^{\rm -1}$)} & \colhead{(keV)} & \colhead{} \\
\colhead{(1)} & \colhead{(2)} & \colhead{(3)} & \colhead{(4)} & \colhead{(5)} & \colhead{(6)} & \colhead{(7)} 
}
\startdata
IRAS F01364--1042 & 1.28 & 1 & 40.20 $\pm$ 0.65 & 42.47 $\pm$ 0.39 & 0.22 $\pm$ 0.50 & 46.65/46 \\
ESO 255-IG7 & 0.82 & 1 & 41.13 $\pm$ 0.05 & 43.42 $\pm$ 0.10 & 0.11 $\pm$ 0.02 & 62.26/51 \\
ESO 60-IG16 & 1.01 & 2 & 40.79 $\pm$ 0.15 & 42.47 $\pm$ 0.35 & 0.24 & 51.77/41 \\
IRAS 09022--3615 & 1.42 & 4 & 41.44 $\pm$ 0.08 & 43.69 $\pm$ 0.41 & 0.17 $\pm$ 0.06 & 34.25/23 \\
IRAS F09111--1007 & 1.06 & 5 & 41.03 $\pm$ 0.18 & 42.42 $\pm$ 0.36 & 0.24 & 54.83/57 \\
ESO 374-IG32 & 0.97 & 2 & 40.54 $\pm$ 0.65 & 41.81 $\pm$ 0.37 & 0.24 & 26.14/32 \\
VV 340 N & 1.06 & 3 & 41.24 $\pm$ 0.03 & 42.16 $\pm$ 0.09 & 0.71 & 88.73/42 \\
IRAS F17132+5313 & 0.98 & 1 & 41.39 $\pm$ 0.05 & 43.31 $\pm$ 0.14 & 0.18 $\pm$ 0.02 & 37.74/31 \\
ESO 593-IG8 & 0.89 & 1 & 41.47 $\pm$ 0.06 & 43.26 $\pm$ 0.15 & 0.17 $\pm$ 0.03 & 34.47/37 \\
IRAS F19297--0406 & 1.83 & 2 & 41.45 $\pm$ 0.14 & 43.48 $\pm$ 0.15 & 0.24 & 25.71/11 \\
CGCG 448-020 & 0.66 & 2 & 41.21 $\pm$ 0.06 & 42.25 $\pm$ 0.14 & 0.24 & 49.11/55 \\
IRAS F22491--1808 & 0.88 & 1 & 40.87 $\pm$ 0.28 & 42.71 $\pm$ 0.54 & 0.19 $\pm$ 0.17 & 11.79/7 \\
ESO 77-IG14 & 1.16 & 2 & 40.87 $\pm$ 0.07 & 42.50 $\pm$ 0.26 & 0.24 & 62.44/50 \\
IRAS F05189-2524 & 1.13 & 5 & 41.17 $\pm$ 0.06 & 42.75 $\pm$ 0.17 & 0.24 & 411.34/378 \\
UGC 4881 & 0.88 & 4 & 40.90 $\pm$ 0.06 & 43.18 $\pm$ 0.10 & 0.14 $\pm$ 0.02 & 56.1/48 \\
UGC 5101 & 1.19 & 1 & 40.95 $\pm$ 0.05 & 43.53 $\pm$ 0.11 & 0.10 $\pm$ 0.02 & 169.97/146 \\
IRAS F10565+2448 & 1.33 & 2 & 41.09 $\pm$ 0.05 & 42.82 $\pm$ 0.32 & 0.24 & 104.17/64 \\
NGC 3690 & 0.70 & 1 & 41.32 $\pm$ 0.01 & 42.60 $\pm$ 0.17 & 0.21 $\pm$ 0.01 & 818.19/510 \\
IRAS F12112+0305 & 1.02 & 1 & 40.81 $\pm$ 0.26 & 42.95 $\pm$ 0.63 & 0.17 $\pm$ 0.08 & 10.85/8 \\
IRAS F14348-1447 & 1.00 & 4 & 41.62 $\pm$ 0.07 & 43.20 $\pm$ 0.16 & 0.22 $\pm$ 0.07 & 26.75/18 \\
VV 705 & 0.82 & 4 & 40.97 $\pm$ 0.07 & 42.72 $\pm$ 0.22 & 0.20 $\pm$ 0.04 & 58.5/47 \\
IRAS F15250+3608 & 0.91 & 1 & 41.26 $\pm$ 0.10 & 43.48 $\pm$ 0.11 & 0.13 $\pm$ 0.08 & 51.65/36 \\
UGC 9913 & 1.41 & 2 & 40.91 $\pm$ 0.01 & 42.83 $\pm$ 0.08 & 0.24 & 1190.94/519 \\
NGC 6240 & 0.94 & 2 & 41.86 $\pm$ 0.01 & 43.32 $\pm$ 0.13 & 0.24 & 1397.98/522 \\
IRAS F17207-0014 & 2.15 & 2 & 41.29 $\pm$ 0.04 & 43.36 $\pm$ 0.33 & 0.24 & 240.44/188 \\
ESO 286-IG19 & 0.84 & 4 & 41.14 $\pm$ 0.04 & 42.78 $\pm$ 0.21 & 0.19 & 183.51/168 \\
ESO 148-IG2 & 0.82 & 1 & 41.28 $\pm$ 0.06 & 42.83 $\pm$ 0.23 & 0.18 $\pm$ 0.04 & 549.96/518 \\
UGC 08387 & 0.90 & 1 & 40.43 $\pm$ 0.08 & 42.47 $\pm$ 0.20 & 0.16 $\pm$ 0.08 & 49.65/33 \\
CGCG 436--030 & 0.74 & 1 & 40.76 $\pm$ 0.10 & 42.95 $\pm$ 0.23 & 0.13 $\pm$ 0.04 & 77.04/73 \\
NGC 0695 & 0.77 & 6 & 41.32 $\pm$ 0.04 & 42.00 $\pm$ 0.06 & 0.71 & 114.24/87 \\
CGCG 043--099 & 0.88 & 2 & 40.63 $\pm$ 0.15 & 42.10 $\pm$ 0.35 & 0.24 & 41.36/27 \\
MCG+07--23--019 & 0.65 & 1 & 41.04 $\pm$ 0.03 & 42.53 $\pm$ 0.12 & 0.18 $\pm$ 0.03 & 227.46/191 \\
NGC 6670 (E) & 0.84 & 2 & 40.38 $\pm$ 0.09 & 41.64 $\pm$ 0.63 & 0.09 $\pm$ 0.13 & 4.9/3 \\
NGC 6670 (W) & 0.97 & 2 & 40.74 $\pm$ 0.06 & 42.18 $\pm$ 0.43 & 0.24 & 21.61/16 \\
NGC 5331 (N) & 0.51 & 2 & 40.57 $\pm$ 0.21 & 41.32 $\pm$ 0.74 & 0.24 & 14.03/4 \\
NGC 5331 (S) & 0.90 & 2 & 41.36 $\pm$ 0.59 & 41.96 $\pm$ 0.49 & 0.24 & 15.72/6 \\
IC 2810 (NW) & 0.85 & 5 & 40.59 $\pm$ 0.14 & 41.80 $\pm$ 0.35 & 0.24 & 24.87/19 \\
NGC 3256 & 0.86 & 1 & 41.10 $\pm$ 0.01 & 42.62 $\pm$ 0.04 & 0.20 $\pm$ 0.01 & 604.17/520 \\
IRAS F16164--0746 & 0.97 & 1 & 40.06 $\pm$ 0.68 & 42.04 $\pm$ 0.46 & 0.11 $\pm$ 0.62 & 93.94/74 \\
IC 4687 & 0.84 & 7 & 40.72 $\pm$ 0.04 & 42.53 $\pm$ 0.15 & 1.15/0.13 & 28.25/28 \\
IC 4689 & 0.81 & 5 & 40.61 $\pm$ 0.06 & 41.67 $\pm$ 0.25 & 0.24 & 25.53/21 \\
IC 5298 & 0.92 & 2 & 40.57 $\pm$ 0.11 & 41.88 $\pm$ 0.21 & 0.24 & 99.43/68 \\
NGC 6090 & 0.48 & 4 & 40.99 $\pm$ 0.06 & 42.29 $\pm$ 0.31 & 0.12 $\pm$ 0.02 & 114.17/104 \\
NGC 5256 & 0.62 & 2 & 41.52 $\pm$ 0.02 & 42.64 $\pm$ 0.16 & 0.24 & 289.68/170 \\
IRAS F03359+1523 & 0.97 & 5 & 40.85 $\pm$ 0.16 & 41.79 $\pm$ 0.44 & 0.24 & 58.42/56 \\
ESO 550--IG025 & 0.80 & 1 & 40.57 $\pm$ 0.61 & 42.27 $\pm$ 0.12 & 0.18 $\pm$ 0.14 & 49.51/45 \\
NGC 0034 & 0.67 & 1 & 40.59 $\pm$ 0.07 & 41.80 $\pm$ 0.28 & 0.22 $\pm$ 0.05 & 67.16/57 \\
MCG+12--02--001 & 1.38 & 1 & 40.59 $\pm$ 0.08 & 42.26 $\pm$ 0.16 & 0.23 $\pm$ 0.12 & 92.7/79 \\
ESO 440--IG058 (S) & 1.07 & 2 & 40.54 $\pm$ 0.06 & 42.17 $\pm$ 0.29 & 0.24 & 16.37/10 \\
NGC 7130 & 0.63 & 1 & 40.79 $\pm$ 0.02 & 42.18 $\pm$ 0.29 & 0.18 $\pm$ 0.01 & 241.83/206 \\
NGC 7771 & 0.83 & 2 & 40.79 $\pm$ 0.02 & 42.11 $\pm$ 0.34 & 0.24 & 138.88/109 \\
NGC 7770 & 0.51 & 1 & 40.25 $\pm$ 0.07 & 41.23 $\pm$ 0.43 & 0.20 $\pm$ 0.05 & 50.43/42 \\
NGC 7592 (E) & 0.36 & 2 & 40.56 $\pm$ 0.05 & 41.11 $\pm$ 0.21 & 0.24 & 18.64/12 \\
NGC 7592 (W) & 0.78 & 5 & 40.28 $\pm$ 0.07 & 41.60 $\pm$ 0.18 & 0.24 & 11.3/7 \\
NGC 6285 & 0.56 & 1 & 40.19 $\pm$ 0.10 & 41.34 $\pm$ 0.21 & 0.18 $\pm$ 0.07 & 40.31/31 \\
NGC 6286 & 0.83 & 3 & 40.99 $\pm$ 0.03 & 41.78 $\pm$ 0.04 & 0.71 & 170.16/104 \\
NGC 4922 (N) & 0.99 & 1 & 40.14 $\pm$ 0.12 & 42.09 $\pm$ 0.62 & 0.15 $\pm$ 0.10 & 3.39/3 \\
NGC 3110 & 0.61 & 1 & 40.84 $\pm$ 0.05 & 41.91 $\pm$ 0.42 & 0.22 $\pm$ 0.04 & 143.11/114 \\
NGC 0232 & 0.91 & 5 & 40.51 $\pm$ 0.05 & 42.01 $\pm$ 0.28 & 0.24 & 70.54/52 \\
MCG+08--18--013 (E) & 0.77 & 5 & 40.51 $\pm$ 0.09 & 41.80 $\pm$ 0.46 & 0.24 & 17.7/16 \\
CGCG 049--057 & 1.61 & 4 & 39.61 $\pm$ 0.31 & 41.07 $\pm$ 0.52 & 0.26 $\pm$ 1.06 & 44.51/41 \\
NGC 1068 & 0.69 & 1 & 41.10 $\pm$ 0.01 & 42.55 $\pm$ 0.11 & 0.12 $\pm$ 0.01 & 4796.17/499 \\
UGC 02238 & 0.98 & 6 & 40.66 $\pm$ 0.08 & 41.47 $\pm$ 0.30 & 0.71 & 62.08/54 \\
MCG--03--34--064 & 0.83 & 2 & 40.57 $\pm$ 0.06 & 41.76 $\pm$ 0.20 & 0.24 & 274.51/191 \\
ESO 350--IG038 & 0.41 & 1 & 40.58 $\pm$ 0.04 & 41.35 $\pm$ 0.18 & 0.21 $\pm$ 0.02 & 423.51/415 \\
MCG--01--60--022 & 0.61 & 1 & 40.61 $\pm$ 0.09 & 42.21 $\pm$ 0.12 & 0.14 $\pm$ 0.05 & 220.8/160 \\
NGC 5135 & 0.67 & 4 & 41.03 $\pm$ 0.01 & 42.23 $\pm$ 0.04 & 0.23 $\pm$ 0.01 & 304.56/255 \\
IC 5179 & 0.60 & 4 & 40.58 $\pm$ 0.04 & 41.78 $\pm$ 0.06 & 0.20 $\pm$ 0.03 & 200.72/168 \\
CGCG 465--012 & 1.11 & 3 & 40.65 $\pm$ 0.09 & 41.49 $\pm$ 0.13 & 0.71 & 56.78/59 \\
MCG-02-33-098-9 & 0.73 & 2 & 39.96 $\pm$ 0.14 & 41.02 $\pm$ 0.50 & 0.71 & 39.76/27 \\
ESO 343--IG013 & 0.75 & 4 & 40.44 $\pm$ 0.07 & 41.92 $\pm$ 0.50 & 0.21 $\pm$ 0.05 & 37.06/35 \\
NGC 2146 & 1.07 & 1 & 40.18 $\pm$ 0.01 & 42.01 $\pm$ 0.09 & 0.19 $\pm$ 0.01 & 918.33/520 \\
NGC 5653 & 0.60 & 1 & 40.37 $\pm$ 0.06 & 42.02 $\pm$ 0.09 & 0.13 $\pm$ 0.03 & 170.28/169 \\
NGC 0023 & 0.64 & 4 & 40.71 $\pm$ 0.06 & 41.60 $\pm$ 0.20 & 0.32 $\pm$ 0.07 & 227.87/205 \\
NGC 7552 & 0.66 & 1 & 40.43 $\pm$ 0.01 & 41.58 $\pm$ 0.23 & 0.23 $\pm$ 0.01 & 589.54/519 \\
NGC 1961 & 0.51 & 1 & 40.75 $\pm$ 0.03 & 41.41 $\pm$ 0.16 & 0.29 $\pm$ 0.05 & 220.48/177 \\
NGC 1365 & 0.65 & 1 & 40.62 $\pm$ 0.01 & 41.93 $\pm$ 0.01 & 0.20 $\pm$ 0.01 & 1371.01/417 \\
NGC 3221 & 0.70 & 4 & 40.36 $\pm$ 0.63 & 41.46 $\pm$ 0.24 & 0.24 $\pm$ 0.07 & 269.53/235 \\
\enddata
\tablecomments{Column (2): intrinsic hydrogen column density. Column (3): best-fitting models for the sample galaxies, as described in Section~\ref{sec:Fitting}. Column (4): apparent luminosity of the hot gas emission in the 0.5$-$2 keV band corrected for Galactic absorption. Column (5): intrinsic luminosity in the 0.5$-$2 keV band corrected for intrinsic absorption. Column (6): temperature of the plasma component. IC 4687 has a low-temperature plasma component of 0.13 $\pm$ 0.03 keV and a high-temperature plasma component of 1.15 $\pm$ 0.26 keV. }
\end{deluxetable*}

\end{document}